\documentstyle[12pt]{article}
\include{epsf}

\newlength{\extralineskip}

\newcommand{\beq}{\begin{equation}}
\newcommand{\eeq}{\end{equation}}
\newcommand{\bd}{\begin{displaymath}}
\newcommand{\ed}{\end{displaymath}}
\def\bea{\begin{eqnarray}}
\def\eea{\end{eqnarray}}

\def\ba{\beq\new\begin{array}{c}}
\def\ea{\end{array}\eeq}

\def\lie{{\cal L}}

\def\inbar{\,\vrule height1.5ex width.4pt depth0pt}
\def\IC{\relax\hbox{$\inbar\kern-.3em{\rm C}$}}
\def\IR{\relax{\rm I\kern-.18em R}}
\def\IZ{{{\rm Z}\!\!{\rm Z}}}

\def\tr{{\rm tr}}
\def\e{~{\rm e}}
\makeatletter
\newdimen\normalarrayskip              
\newdimen\minarrayskip                 
\normalarrayskip\baselineskip
\minarrayskip\jot
\newif\ifold             \oldtrue            \def\new{\oldfalse}
\def\arraymode{\ifold\relax\else\displaystyle\fi} 
\def\@arrayskip{\ifold\baselineskip\z@\lineskip\z@
     \else
     \baselineskip\minarrayskip\lineskip2\minarrayskip\fi}
\def\@arrayclassz{\ifcase \@lastchclass \@acolampacol \or
\@ampacol \or \or \or \@addamp \or
   \@acolampacol \or \@firstampfalse \@acol \fi
\edef\@preamble{\@preamble
  \ifcase \@chnum
     \hfil$\relax\arraymode\@sharp$\hfil
     \or $\relax\arraymode\@sharp$\hfil
     \or \hfil$\relax\arraymode\@sharp$\fi}}
\def\@array[#1]#2{\setbox\@arstrutbox=\hbox{\vrule
     height\arraystretch \ht\strutbox
     depth\arraystretch \dp\strutbox
     width\z@}\@mkpream{#2}\edef\@preamble{\halign \noexpand\@halignto
\bgroup \tabskip\z@ \@arstrut \@preamble \tabskip\z@ \cr}%
\let\@startpbox\@@startpbox \let\@endpbox\@@endpbox
  \if #1t\vtop \else \if#1b\vbox \else \vcenter \fi\fi
  \bgroup \let\par\relax
  \let\@sharp##\let\protect\relax
  \@arrayskip\@preamble}

\begin{document}

\begin{titlepage}
\setcounter{footnote}0
\rightline{\baselineskip=12pt\vbox{\halign{&#\hfil\cr
&hep-th/96xxxxx &\cr
{   }&\cr &Revised Version {   }&\cr &\today\cr}}}
\vspace{0.5in}
\begin{center}
{\Large\bf Vacuum Structure
and $\theta$ States of Adjoint QCD in Two Dimensions}\\
\medskip
\vskip0.5in
\baselineskip=12pt

\normalsize {\bf Lori D. Paniak}\footnote{E-mail: paniak@physics.ubc.ca},
{\bf Gordon W. Semenoff}\footnote{E-mail: semenoff@physics.ubc.ca}
and {\bf Ariel R. Zhitnitsky}\footnote{E-mail: arz@physics.ubc.ca}
\medskip

\baselineskip=12pt

{\it Department of Physics and Astronomy, University of British Columbia\\
Vancouver, British Columbia, Canada V6T 1Z1}

\end{center}
\vskip1.5in

\begin{abstract}
\baselineskip=12pt
We address the issue of topological angles in the context of two dimensional
SU(N) Yang-Mills theory coupled to massive fermions in the adjoint representation.
Classification of the resulting multiplicity of vacua is carried out
in terms of asymptotic fundamental Wilson
loops, or equivalently, charges at the boundary of the world.
We explicitly demonstrate
that the multiplicity of vacuum states is  equal to   $N$ for $SU(N)$ gauge group.
Different worlds of the theory
are classified by the integer number $k=0,1,...N-1$
(superselection rules) which plays an analogous
role to the $\theta$ parameter in QCD. We study
the physical properties of these unconnected worlds as a function of $k$.
We achieve this by using two completely independent approaches: First,
we apply the well known machinery of the loop calculus in order
to calculate the effective string tensions in the theory as function of $k$.
The second way of doing the same physics
is the standard particle/field theoretic calculation for the
binding potential of a pair of infinitely massive fermions. We also
calculate the vacuum energy as function of $k$ 

\end{abstract}
\end{titlepage}
\section{Introduction}

The existence of topological parameters in certain quantum field theories,
such as the $\theta$ - angle of quantum chromodynamics (QCD)\cite{Gross} or
 Schwinger model (\cite{coleman}, \cite{suss}) are
interesting possibilities which can have a profound effect on the
physical properties of the models.
As is known, many problems such as the  $U(1) $ problem,
chiral symmetry breaking phenomenon, confinement and
multiplicity of vacuum states are related
to each other and to $\theta$ dependence.
Therefore, in spite of the fact that
we live in a certain vacuum state (superselection rule),
the variation of physical values with  $\theta$
is an extremely important characteristic of the theory.
Unfortunately we do not know the nature of such a  dependence in a 
physically interesting
theory like four-dimensional QCD which appears too complicated to deal with.
The $\theta$ dependence in Schwinger model can be found exactly,
however this model is too trivial to share some important
QCD- features. 

An example of a theory which admits such angles and
shares many features of four dimensional gauge theories
is 1+1-dimensional QCD with adjoint matter as was pointed out many years ago by Witten\cite{witten}.
This  model demonstrates a complex spectrum
\cite{kutkleb},
a Hagedorn transition at finite temperature (\cite{kutkleb}, \cite{arz},
\cite{zarembo}) and is a testing ground for answering questions
about the phenomena of screening and confinement \cite{gsmk},
and many other questions, some of which were pointed out above.
Studying these features in the context of a $1+1$-dimensional field theory
is desirable since a lack of
transverse degrees of freedom in two dimensional space-time,
and consequently dynamical gluons, makes
explicit calculations possible.

In order to gain a full comprehension of these features it is
important to understand the vacuum structure of the theory.
In particular there are many outstanding
issues pertaining to the link between the existence of topological
parameters, multiple vacua,
fermion condensates and instantons.  For example if one considers
the bosonized version of the theory there appear to be fermion condensates
for any SU(N) gauge group and such a condensation
is also supported by very general quark-hadron
duality arguments \cite{arz}.
However the instanton calculations only show the
existence of condensates when $N=2$ \cite{smilga}.
The path to resolving this paradox was defined in \cite{shifman} where
the quantum mechanics of the vacua in the finite volume limit 
for $N=2,3$ was investigated (see also \cite{pinsky} for the finite volume
calculation with $N=2$).
Here our goal is to proceed with this investigation of the
properties of the multiple vacua in this model for arbitrary $N$
via a different route with a continuum description from the 
outset.  We will consider the  limiting case of
infinite fermion mass and use the known methods of Wilson loop calculations
to gain insight into the interactions of a pair of adjoint particles
as a function of topological parameters.  In order to make definite contact
with the particle interpretation of these loop calculations we also
consider  the theory through explicit calculation of the
effective Hamiltonian in the limit of large fermion mass.

The standard method of classifying the multiplicity of vacua
in a particular gauge theory with adjoint matter hinges on
identifying the effective gauge group. Here, since gauge
transformations operate by adjoint action on all fields,
the true gauge group is the quotient of the gauge group and its
center.  This quotient is multiply connected.  For simply connected
semi-simple gauge group $G$ with center $Z$,
\beq
\pi_1( G/Z)~=~ \pi_0(Z)~=~ Z
\eeq
This gives a classification of gauge fields which are constrained to
be flat connections at infinity.  In that case
\beq
\lim_{ |x| \rightarrow \infty}~A_\mu(x)~= ig^{\dagger}(x)\nabla_\mu
g(x)
\eeq
Where $g(x)$ is a mapping of the circle at infinity to the gauge group
$G/Z$. Since $G/Z$ is a symmetry of the Hamiltonian,
we expect that all physical states carry a representation of $\pi_1(G/Z)$.
In the case where the center of the group is Abelian
all of its irreducible representations are one dimensional and further, when
$Z \sim \IZ_N$, as in the case of SU(N), we are lead to a classification of all physical states in
terms of a single integer parameter $k$.
If ${\cal Z }$ is a generator of $Z$ and $| \psi >$ is a physical state we have
\beq
{\cal Z} |\psi > = \e^{2 i \pi k /N} | \psi >
\eeq

An alternate method of classifying the vacua in such a model
follows the example of the Schwinger model (\cite{coleman}, \cite{suss})
and involves coupling the gauge fields to external static charges
which reside at the boundary of the world. This method was used
by Witten \cite{witten} to identify $\theta$-vacua in two dimensional
non-Abelian field theories. Here we will further develop these ideas
and obtain quantitative results using Wilson loop calculations.
First though, let us review this classification scheme.

As pointed out by Coleman in the case of the massive Schwinger
model \cite{coleman}, one introduces the parameter $\theta$ as
the strength of a fractional charge $e \theta /2 \pi$ at the right
boundary of the world and an opposite charge at the left boundary of
the world. Hence we see that topological parameters here can be considered
as a form of generalized boundary conditions for the theory.
The most important aspect of this picture is interpreting these
charges in terms of a Wilson loop enclosing the world which  can
be explicitly included in the action of the theory
\bea
Z &=& \int {\cal D} \psi {\cal D} \bar{\psi} {\cal D} A
\exp{(- \int d^2 x ~\lie)} \nonumber \\
& \rightarrow & \int {\cal D} \psi {\cal D} \bar{\psi} {\cal D}
A \exp{(- \int d^2 x ~\lie + \frac{i \theta}{2 \pi}
\oint_{C \rightarrow \infty} dx^\mu A_\mu(x)  )}  \nonumber \\
& = &\int {\cal D} \psi {\cal D} \bar{\psi} {\cal D}
A \exp{(- \int d^2 x ~\lie_{\theta})},~~~~ \lie_{\theta}\equiv
\lie-\frac{i\theta\epsilon_{\mu\nu}F_{\mu\nu}}{4\pi}
\label{abelian}
\eea
Similarly, in the non-Abelian case we consider static colour charges
$T_R$ and $T_{\bar{R} }$ at either end of the world.  Here the $T$'s
are the generators of the colour group in the representation $R$
and its conjugate, respectively.  Unlike the Abelian case though,
we do not have any continuous parameter only the discrete choice
of representation of the boundary charges.
If $\tr_R$
is the trace in the representation $R$ of the gauge group then our theory
is modified \cite{witten} to
\beq
Z \rightarrow \int {\cal D} \psi {\cal D} \bar{\psi} {\cal D} A
\exp{(- \int d^2 x ~\lie )}~
\tr_R P \exp{(i \oint_{C \rightarrow \infty} dx^\mu A_\mu (x) )}
\label{bloop}
\eeq
The number of distinct choices of representation in which to take the
trace of the boundary loop is directly related to the transformation
properties of the loops under $Z$, the center of the gauge group.
We note that unlike the  Abelian case the formula (\ref{bloop})
can not be written in terms of some local lagrangian $\lie_{\theta}$
(\ref{abelian}).

As we have seen, the effective gauge group is divided into $N$ different
classes by modding out the center of SU(N) and so we expect that the
only important feature of any external charges that we may add to
the model are their transformation properties under the center.
In fact we may think that all we need to do is choose representatives
from each class and do our calculations with those.  This is indeed
the case if we take into account the stability of our choice of
external charge.  Since all charges are coupled to dynamical
fermions there is always the possibility of the vacuum producing
adjoint pairs to screen charges whenever this is energetically
advantageous.
Witten pointed out that the only stable vacua are
those for which the boundary charges/loops are taken in one of
the $N$ (antisymmetric) fundamental representations
(including the trivial one) of SU(N), each
of which transform differently under the center.
An explicit example of  unstable representations
will arise when we consider the
external charges in terms of Wilson loops at the boundary of the
world.

Labeling the $N$ different stable vacuum states by the index $k$, we see
the  analysis in terms of Wilson loops is  consistent with the previous
homotopy arguments. While it is an open question if 
this agreement holds for an arbitrary gauge group, we will concentrate
here on the case of adjoint QCD with a
gauge group SU(N) and center $Z \sim \IZ_N$. 
{\bf Our goal is to answer questions
about the physics in each sector of the theory as a function of the discrete
topological angle label $k \in \{0,1, \ldots, N-1 \}$}.

We  will now proceed via two routes, one group theoretical
and the other particle/field theoretical.
First we will introduce the machinery
originally due to Rusakov \cite{rusa} for the calculation of Wilson loop
correlators in pure non-Abelian gauge theories.
Using these techniques we will calculate the effects of fundamental
boundary loops on a world that contains a single adjoint
loop which corresponds to a pair of adjoint fermions in the particle
picture of the problem.
In  terms of physical variables these calculations correspond
to the evaluation of the string tension for the heavy meson
constructed from a pair of adjoint quarks (adjoint internal loop)
in  $k$-th vacuum state ( fundamental boundary loop).

The {\bf same} system can be analyzed by the
standard, pure Hamiltonian, approach. As such
we will introduce the model of a single flavour of adjoint fermions
in two dimensional QCD and discuss the introduction of
the boundary loop (\ref{bloop})
in terms of a modified colour electric field due to charges at the
edges of the world.  Once the model is set we will
look at its quantum behaviour via canonical quantization in the limit of
infinite fermion mass.  Here, as in the massive Schwinger model, the field
theory problem is reduced to one dimensional quantum mechanics and
we show that the number and qualities of bound and bleached states
have a clear interpretation in terms of the previous
Wilson loop calculation.  Finally we
present a discussion and conclusion with ideas for future directions
with this model.

\section{Wilson Correlators in 2D Gauge Theory}

The fundamental gauge invariant object in the study of gauge theories
is the Wilson loop operator which is defined in terms of a path ordered
product of gauge field operators about some closed path (loop) in
space, $C$
\beq
W(C) = \tr_R P \exp{ ( \oint_C dx^\mu A_\mu (x) )}
\eeq
Explicit calculation of the averages of such operators, which form
a natural set of observables in two dimensional Yang-Mills theory,
is an old
topic which arose in the loop formulation of QCD by Makeenko and
Migdal \cite{mak}.
Using Schwinger-Dyson methods it is possible to derive correlators for
arbitrary loop configurations
on the plane \cite{kaz} from a system of partial differential
equations which relate the correlators of
loops with $n$ self-intersections to those with fewer crossings.
While in principle these equations are
soluble in general, we are required to solve an entire system
in order to find an expression for a single Wilson loop average. Another,
group theoretic approach by Rusakov \cite{rusa} developed
later allows straightforward
calculation of correlators for nested loops on the plane
and is more useful for our applications.
Since we have invariance under area preserving diffeomorphisms in
two dimensional Yang-Mills theory we expect that for a series of nested
loops on the plane $C_1 \ldots C_n$  each taken in representation
$R_1 \ldots R_n$, the weight of the entire
configuration is a function only of the areas between successive
loops $S_1 \ldots S_n$.  While the explicit form of the Wilson
correlator $<W(C_1 \ldots C_n)>$ is cumbersome to write in a transparent
fashion, the general
method of calculation follows easily  from generalization of the
basic examples.

In evaluating Wilson correlators there is a
physical picture one can consider due to Minahan and
Polychronakos \cite{poly} (see also \cite{gross} ).
Consider nested loops $\{C_k \}$ as rings around a cylinder,
with periodic space components and time running along the axis of
the cylinder, each associated with a character of the appropriate
representation $\{ \chi_U (R_k ) \}$.
Information is then carried through time -the area
on the cylinder between successive loops $\{ S_k \}$-  via the
propagator $K(R_i,R_j)$  which is the exponential of the
quadratic Casimir of the representation under transport
\beq
K(R_i,R_j) = \delta_{R_i R_j} \exp{( -g^2 /2 ~ C(R_i) S_i )}
\eeq
The interaction of two loops
is due to the property of the group characters
\beq
\chi_U (R_i) \chi_U (R_j) =  \sum_k \chi_U (R_{k})
\eeq
Where $ R_i \otimes R_j = \sum_k R_k $ is the decomposition of the
product representation into irreducible components.
The Wilson correlator is then given as an integral of
the sum over branches of such products of
representations with the relevant propagators
where each term carries a factor for the dimension
$\{ \chi_{U^\dagger}(R_k) d(R_k) \}$ of each representation that
makes it to the end of the cylinder.

For example, we would like to calculate the Wilson correlator of a single
adjoint SU(N) loop which encloses an area $S$ in  Euclidean
two dimensional space-time.  In terms of the previous picture we have a
loop on the cylinder an area S from the future end of the
cylinder which is closed to a point. Hence
\bea
<W(C)> &=& \int dU ~\chi_U (R_A)~ \exp{( -g^2 /2~ C(R_A) S )} ~
\chi_{U^\dagger}(R_k) d(R_k) \\
&= & d(R_A) \exp{( -g^2 /2 ~C(R_A) S )} \nonumber \\
& = & (N^2 -1) \e^{-g^2 N S}
\label{plain}
\eea
where we have used the orthogonality condition
\beq
\int dU ~\chi_U (R_i) \chi_{U^\dagger} (R_j) =
\delta_{R_i R_j}
\eeq

Similarly we can write down the Wilson correlator for a pair of
nested loops each in the N-dimensional representation $(R_F)$ of SU(N).
If the outer loop $C_1$ occupies a total area of $S_1+S_2$ and
the inner $C_2$ loop $S_2$ then
\beq
< W(C_1,C_2) > ~~~~~~~~~~~~~~~~~~~~~~~~~~~~~~~~~~~~~~~~~~~~~~~~~
~~~~~~~~~~~~~~~~~~~~~~~~~~~~~~~~~~~~~~~~~~~~~~~~~~~~~~~~~~~~~~~~~~~
~~~~~~~~~~~~~~~~~
\eeq
\bd
= \int dU \chi_U (R_F)  \exp{( -g^2 /2 ~C(R_F) S_1 )} ~
\chi_U (R_F)~
\sum_k \chi_{U^\dagger}(R_k) d(R_k) \exp{( -g^2 /2~ C(R_k) S_2 )}
\ed
Now we decompose the product of representations
$R_F \otimes R_F =  R_1 \oplus R_2$ and integrate over the group
manifold
\beq
< W(C_1,C_2) >~~~~~~~~~~~~~~~~~~~~~~~~~~~~~~~~~~~~~~~~~~~~~~~~~
~~~~~~~~~~~~~~~~~~~~~~~~~~~~~~~~~~~~~~~~~~~~~~~~~~~~~~~~~~~~~~~~~~~
~~~~~~~~~~~~~~~~~
\eeq
\bea
&=&  \exp{( -g^2 /2 ~C(R_F) S_1 )} [  d(R_1) \exp{( -g^2 /2 ~C(R_1) S_2 )}
+ d(R_2) \exp{( -g^2 /2 ~C(R_2) S_2 )} ]
\nonumber  \\
&= & \frac{N}{2} \e^{ -g^2 \frac{N^2 -1}{2N} (S_1 +S_2) }
[(N+1) \e^{ -g^2\frac{ (N-1)(N+3)}{2N} S_2} +
(N-1) \e^{ -g^2\frac{ (N+1)(N-3)}{2N} S_2}]
\nonumber
\eea

As pointed out in the introduction, we would like to
investigate the effect of external fundamental loops on the  Wilson
correlator of a single adjoint loop in the world.
These fundamentals are antisymmetric combinations of single
boxes in the Young Tableaux notation each transforming distinctly under
an element of the center of SU(N).  In this notation the
$k$-fundamental representation is denoted by a single column of $k$
boxes. In tensor notation, which we will also use, these representations
are given as contractions of the N-dimensional completely anti-symmetric
tensor with $k$ single index fundamental charges
$\epsilon^{i_1 \ldots i_{N-k} m_1 \ldots m_k} u^{i_1} \ldots u^{i_k} $.
Clearly from this last definition $k$ defined modulo N only.
Later we will discuss
the quantum field theory situation with adjoint charges in the limit
of infinite mass and make comparisons with the following calculation.
Precisely we will use the previously introduced machinery
to calculate the Wilson correlator of a k-fundamental loop $C_1$ of
total area $S_1+S_2$ within which lies an adjoint loop
$C_2$ of area $S_2$.
A straightforward generalization of the previous examples leads to
the result
\bea
<W(C_1,C_2)>
=&\frac{ N!}{(N-k)! k!} \e^{-g^2(S_1 + S_2)
\frac{k(N-k)(N+1)}{2N}}
[ 1 + \frac{(N-k-1)(N+1)}{k+1} \e^{ -g^2 S_2 (N-k) } 
\label{static} \\
&+ \frac{ (N+1)(k-1)}{N-k+1} \e^{-g^2 S_2 k}
+\frac{k N(N+2)(N-k)}{(k+1)(N-k+1)} \e^{-g^2 S_2 (N+1)}] \nonumber
\eea
Where we will take the leading factor which is just the contribution of a
k-fundamental loop of total area $S_1+S_2$ to be normalization and
ignore it in any further discussions.

The first interesting observation
to be made about (\ref{static}) is the effect of the addition of a
boundary loop on the number of energetically distinct singlet
configurations. Without an external loop ($k=0$) the adjoint
loop can form only a single stable configuration (\ref{plain}) while
with $k \ne 0$ the system allows up to four distinct configurations.
Also, for $k=N$ we expect the external loop to form a   singlet
itself and not contribute to the effective Wilson correlator
for the adjoint loop.  This expectation is met in (\ref{static})
and we explicitly see that the physics of adjoint loops in this model
depends on $k$ mod N in analogy with the continuous $\theta$-angle
which is periodic in $2 \pi $.

Also of note
is a symmetry under changing the representation of the external
loop by $k \rightarrow N-k$.  Mathematically this is the procedure
of replacing the gauge fields along the path $C_1$ with their conjugates,
which we denote $\bar{C}_1$, and should lead to
changes in (\ref{static}) for loops of arbitrary representation in the world.
The situation here is special though
since we are dealing with $C_2$ in
an adjoint representation which is invariant under charge conjugation,
$C_2 = \bar{C}_2$.
This invariance is a manifestation of the general invariance of
the adjoint representation under transformations in the center of the
gauge group which leads to the phenomena of $\theta$-vacua.
A consequence of this symmetry we see that the vacua corresponding to
$k$ and $N-k$ are degenerate in energy and hence there are
only $(N+2)/2$ and $(N+1)/2$  {\em distinct }
loop interaction configurations
possible for $N$ even and odd, respectively.
However the total number of states is equal to $N$
as was expected.

The usual method for systematically characterizing these configurations
is through the definition of the string tension between the fermions.
Disregarding the leading factor in (\ref{static}) each term  identifies a
different charge configuration where the prefactor denotes the degeneracy
and the exponent, energy. For each configuration we define the string
tension $\sigma$ to be this energy divided by the area of the loop. 
For the case of $k=1$  we find string tensions of 
\beq
0 ,~~~g^2 (N-1),~~~g^2 (N+1) 
\label{strten}
\eeq
for gauge group SU(N).
Later we will compare these string tension calculations with
similar ones derived through a different analysis of the same
physical problem, but first some comments on having a zero string tension.
The vanishing of the string tension means that we have
states which are  color singlets formed from the
  two quarks with zero binding energy. Therefore, those quarks can travel
in any direction and can be separated for
 a long time at  any distance. At the same time this system is a  color singlet state.
Such a behavior corresponds to a quark-antiquark correlation at large distances
and might be a demonstration of a long- distance order.
Therefore, this effect presumably is related somehow to condensation
of the fermion fields which will be discussed elsewhere.
We would like to recall here that similar
behavior was analyzed in a different model: $QED_2$ with $N$ flavours and nonzero
$\theta$ angle,
\cite{swieca}. Namely, it was found that for the
 special values $\theta=\frac{\pi}{N}k ~~k=0,1...N$
the system possesses some ``exotic" massless states
which were called `` screened quarks''. We note that
those $\theta=\frac{\pi}{N}k$ are very similar to our
vacuum states where such `exotic" states are also exist. 
Moreover,  those values are precisely vacuum states
for which the model becomes $P$ and $T$ invariant--
a property which holds in our model too as we will see in section 4.

As mentioned in the introduction, we only consider the external
loops in the $k$-fundamental representations since all other
representations are unstable due to pair production of
dynamical adjoint
charges.  While this can be argued from basic principles
\cite{witten}, here we can demonstrate  this phenomena
explicitly in the formalism
of loop averages.  For simplicity we consider
the gauge group to be SU(2) for
a moment and label the external
representations by their spin $l/2$.  In the same geometry as
(\ref{static}) we find that the correlator of an adjoint loop
in this background is
\beq
<W(C_1,C_2)> = \e^{- g^2 \frac{l(l+2)}{4} (S_1 + S_2)}
[ (l+1) + (l+3) \e^{-g^2 (l+2) S_2} + ( l-1) \e^{ g^2 l S_2}]
\eeq
If we consider the prefactor to be the normalization due to a
single loop of spin $l/2$ and area $S_1 + S_2$, then we find
that for $l \ge 2 $ the last term is divergent for large values of $S_2$.
This configuration has a negative string tension and is associated with
a large amount of energy that could be used to pair creation had this
calculation involved dynamical fermions.  As it stands, we see that
in order to obtain meaningful results with loop calculations it is
necessary to consider only stable configurations of the full
model.

In passing it is interesting to note that these calculations
lead to  non-perturbative results for observables pure
gluo-dynamics in the presence of topological angles.  For example
we consider the world encircled by a single Wilson loop in one of the
$k$-fundamental representations.  By (\ref{bloop}) the expectation
value of such a loop in nothing more than the partition function of the
problem in the  $k-th$ vacuum state
and hence we can calculate the vacuum energy as a function
of the external parameter $k$.
Since our original (Euclidean) action is
$\frac{1}{4 g^2} \int F^a_{\mu \nu}(x) F^{a \mu \nu}(x) d^2 x $,
differentiating the expectation value of the loop with respect
to $1/g^2$ we find
\beq
\label{Eucl1}
-\frac{1}{4} \int d^2x \langle k|F^a_{\mu \nu}(x) F^{a \mu \nu}(x) |k\rangle =
V g^4 \frac{ k (N-k) (N+1)}{2N}
\eeq
And the {\it Euclidean} vacuum energy $T_{00}$   in volume $V$ is simply
\beq
\label{Eucl}
T_{00} /V = - g^2 \frac{ k (N-k) (N+1)}{2N}
\eeq
A few remarks are in order. First,
 the sign (\ref{Eucl})  corresponds to the positive
vacuum energy for the original Minkowski theory
for all $k$ except $k=0$ and $ k=N$ where the vacuum energy equals zero.
We note also that our calculation of the  vacuum energy corresponds
to the nonperturbative part   with subtracted perturbative pieces
: $E_{vac}=E^{total}_{vac}-E^{pert}_{vac}$ (see for example \cite{Shifman}).
Therefore, such a vacuum expectation value could have,
in principle, any sign, positive or negative.
For example, the positiveness of the operator
(in Euclidean space)  $F^a_{\mu \nu}(x) F^{a \mu \nu}(x)$
 does not guarantee the positiveness of its vacuum expectation value.
  In particular, in 4D QCD one expects a negative  vacuum energy 
while in the 2D $CP^{N-1}$ model for large $N$
the vacuum energy is positive \cite{Shifman}. 
In $QCD_2(N=\infty)$ with massless fermions in the
fundamental representation (t'Hoft model) the vacuum energy has negative sign again
\cite{AZ} (all these results quoted for the Minkowski space). We also note that
our normalization corresponds to the case
when vacuum energy iz zero when   no external Wilson loop
is inserted. Therefore, we essentially study an  energy difference between different
$|k\rangle$ vacuum states. Finally we point out 
the dependence on $N$  in (\ref{Eucl})
comes in the standard combination $g^2 N$ for large $N$
with an extra factor $\frac{k}{N}$.
The appearance of $k$ in this way
is reminiscent of $\theta$ dependence in the large $N$ limit of
gauge theories where we expect the combination $\frac{\theta}{N}$ to 
occur  \cite{Witten1}.

\section{2D adjoint QCD $(m \rightarrow \infty)$ }

Now we change our perspective on the problem at hand and
consider the particle/field
theoretic  model of  adjoint fermions coupled to an SU(N) gauge
theory.  We will show that this model, in the limit of infinite
fermion mass, is nothing more than a different formulation of
the Wilson loop problem for the adjoint loop in a $k$-fundamental
background which we have been considering.
In the past calculations have been
based mainly on bosonization of the problem and utilized
the technology of light-cone coordinates
\cite{kutkleb}, but
here we will take a much simpler approach. We will follow
closely  Coleman's analysis of topological angles in the massive
Schwinger model \cite{coleman} and consider the case of extremely massive
fermions.  As we will see, in this limit the field theory problem is
reduced to that of one dimensional quantum mechanics.
Using such a simplified version of $QCD_2$ is sufficient to carry out our
analysis of the vacuum structure and
general properties of the spectrum, each as a function of
vacuum angle $k$. 

We begin with the Lagrangian density of a single flavour of
adjoint fermion minimally
coupled to a colour electric field in two dimensions
\beq
\lie = \tr[ -\frac{1}{2} F^{\mu \nu} F_{\mu \nu} + \bar{ \psi} ( i D -m)
\psi]
\eeq
Where $iD_\mu \psi = i\partial_\mu \psi - g [A_\mu , \psi]$ and the
fields are expanded on a basis of hermitean matrices $T^a$
\beq
\psi = T^a \psi^a~~~,~~~ F_{\mu \nu} = F^a_{\mu \nu} T^a
\eeq
which are normalized
$ tr T^a T^b = \frac{1}{2} \delta^{a b}$.
Here we use the $\gamma$ matrices
\beq
\gamma^0 = \sigma_2 ~~~,~~~\gamma^1 = i \sigma_1 ~~~,~~~
\gamma^5 = \gamma^0 \gamma^1
\eeq
Hence in more explicit notation where $f^{abc}$ are the structure constants
of the colour group
\beq
\lie =  -\frac{1}{4} F^{a \mu \nu} F_{\mu \nu}^a +
\frac{1}{2}[ \psi^{a \dagger}(  i \gamma^0 \gamma^\mu \partial_\mu  - \gamma^0 m)
\psi^a] + \frac{ g }{2} \psi^{a \dagger} \gamma^0 \gamma^\mu \psi^b f^{abc} A^c_\mu
\eeq
From here we can read off the colour-electric current
\beq
J^a _\mu = \frac{1}{2} \psi^{b \dagger} \gamma_\mu \psi^c f^{b a c}
\eeq
There are gauge degrees of freedom in this problem which we would
like to eliminate and in order to do so we make the
most convenient choice, the  (axial, Coulomb) gauge $A_1 = 0$.
Under this choice the
field strength tensor becomes  $F^a_{0 1} = - \partial_1 A^a _0
= E^a$ and if
we now consider the equations of motion for the gauge field we find
the two dimensional version of Gauss' Law
\beq
\partial_1 E^a = - g J^a_{0 T}
\eeq
Where $J^a_{0 T} $ represents the total colour charge distribution of the system.
The general solution of this equation is
\beq
E^a(x)  = - \frac{g}{2} \int dy ~\epsilon(x - y) J^a_{0 T}(y)
\eeq
As described in the introduction, the
inclusion of loops of gauge fields in the action explicitly
brings topological angles into the model and leads to a background
colour-electric field.  Here we will generate this background field as
was done in  \cite{coleman} and \cite{witten} by
introducing external charges at the edges
$( x = \pm L)$ of our world.
This involves taking the total colour charge
to be the usual charge associated with the fermions $J^a_0$ plus a boundary
contribution
\beq
J^a_{\mbox{ext}} (x) = \delta( x+ L) T^a_{\bar{F}} +
\delta( x- L) T^a_{F}
\eeq
where the operators $T^a_{\bar{F}},T^a_{F}$ act on independent subspaces
and can be physically regarded as being a static charge of representation $F
(\bar{F})$ at $x = L(-L)$.  We are now interested in the energy of such a
configuration
\beq
\int~E^a(x)E^a(x) ~dx = -\frac{g^2}{2} \int |y-z|
J^a_{0 T}(y) J^a_{0 T}(z) ~dy dz
\eeq
In order to have a finite energy associated with a charge configuration
it is necessary that we only deal with charge singlets.  Here this is
defined by the statement that the total charge operator
$Q^a_T$ annihilates any physical
states where
\beq
Q^a_T = \int~J^a_{0 T}(x) ~ dx =
\int~J^a_{0 }(x)~dx + T^a_{\bar{F}} + T^a_{F}
\label{qt}
\eeq
With this restriction we obtain
\bea
\int~E^a(x)E^a(x) ~dx & =& -\frac{g^2}{2} \int |y-z|
J^a_{0 }(y) J^a_{0 }(z) ~dy dz
- g^2 \int~dy~y J^a_0(y) ( T^a_{\bar{F}} - T^a_{F})
\label{cenergy} \\
&&+ 2 L g^2 C_2(F) \nonumber
\eea
We note that the last term is simply the contribution that one would
expect to the energy of a system consisting of two static, conjugate
charges separated by a distance of $2L$. This is consistent with the exponent
of the calculation for a single Wilson loop as we have seen,
for example, in (\ref{plain}).

Calculating the conjugate momentum and adding the energy of the
colour-electric field (\ref{cenergy})
we arrive at the Hamiltonian formulation
of the problem
\bea
H &= & \frac{1}{2} \int~dx~ \psi^{a \dagger}(x)
(  i \gamma_0 \gamma_1 \partial_1 +  \gamma_0 m) \psi^a(x) +\int~E^a(x)E^a(x) ~dx\\
&& = \frac{1}{2} \int~dx ~\psi^{a \dagger}(x)
(  i \gamma_0 \gamma_1 \partial_1 +  \gamma_0 m) \psi^a(x)
\label{hamil} \\
&&  -\frac{g^2}{2} \int |y-z|
J^a_{0 }(y) J^a_{0 }(z) ~dy dz
- g^2 \int~dy~y J^a_0(y) ( T^a_{\bar{F}} - T^a_{F}) \nonumber
\eea
where we have dropped an infinite constant.

\section{Quantization}

We will now proceed to determine the quantum behavior of (\ref{hamil})
in the limit of large fermion mass. The first thing to note is
the Hamiltonian we have
constructed contains particle number changing terms which in general require
a full relativistic field theory treatment.  Considerable simplification
can be realized though if we restrict ourselves to a two-particle subspace
and consider number changing interactions as a perturbation.  In fact
the limit of large fermion mass will be seen to coincide with the weak
coupling limit of the original theory and as demonstrated by the
analogous calculation of Coleman for the massive Schwinger model
\cite{coleman} we can systematically find the
effective Hamiltonian to zeroth
order in $\hbar$. In fact the only (minor) difference between
our calculation and the previous results stems from the fact we use
a Majoranna field to represent our adjoint fermions.
We use the following  mode decomposition of the field
\beq
\psi^a(x) = \int \frac{dp}{\sqrt{2 \pi} } [ b^a(p)~u(p)~\e^{-ip.x} +
 b^{a \dagger}(p)~v(p)~\e^{ip.x}]
\eeq
where equal time anti-commutators satisfy
\beq
\{b^r(p), b^{s}(q) \} = \{b^{r \dagger}(p), b^{s \dagger}(q) \} = 0~,~
 \{b^r(p), b^{s \dagger}(q) \} = \delta^{rs} \delta(p-q)
\eeq
and we define the spinors in our choice  of $\gamma$ matrices
\beq
u = v^* = \frac{i}{\sqrt{2 E(p) (E(p) + p) } } \left( \begin{array}{c} -i m \\
 E(p) + p
\end{array} \right)
\eeq
where $E(p) = \sqrt{ p^2 + m^2} $

The construction of states which are annihilated by the total charge
operator (\ref{qt}) is straightforward and involves considering the
possible contractions of indices in the product state of the fermion,
external charge and adjoint external charge state vectors.  To be
explicit we have in the most general case
\beq
(T^{\mu})_{ab} (T^{\nu})_{cd}
b^{\mu \dagger}(u)b^{\nu \dagger}(v)|0 > \otimes
\epsilon^{i_1 \ldots i_{N-k}m_1 \ldots m_k} |m_1> \ldots |m_k> \otimes
\epsilon^{j_1 \ldots j_{N-k}n_1 \ldots n_k} |\bar{n}_1> \ldots |\bar{n}_k>
\eeq
where the first factor is the configuration of charges for the fermions
with $T^\mu$ a generator of the colour Lie algebra.  The last two factors
represent the external charges at the boundary of the world, the second for
the the state with external
charge $k$ and the third for its adjoint.  All indices run over the dimension
of the defining representation of SU(N) and
$\epsilon^{i_1 \ldots i_N}$ is the completely anti-symmetric invariant
tensor in $N$ dimensions.
One can easily convince oneself that there are only four independent
ways to contract the indices $\{ a,b,c,d,i_1,\ldots,i_{N-k},
j_1,\ldots,j_{N-k} \}$ which are
\bea
\label{1}
|1> & \sim & b^{\mu \dagger}(u)b^{\mu \dagger}(v)|0 > \\
& \otimes &
\epsilon^{i_1 \ldots i_{N-k}m_1 \ldots m_k} |m_1> \ldots |m_k> \otimes
\epsilon^{i_1 \ldots i_{N-k}n_1 \ldots n_k} |\bar{n}_1> \ldots |\bar{n}_k>
\nonumber
\eea
\bea
\label{2}
|2> & \sim &(T^{\mu}T^{\nu})_{i_1 j_1}
b^{\mu \dagger}(u)b^{\nu \dagger}(v)|0 > \\
& \otimes &
\epsilon^{i_1 i_2 \ldots i_{N-k}m_1 \ldots m_k} |m_1> \ldots |m_k> \otimes
\epsilon^{j_1 i_2\ldots i_{N-k}n_1 \ldots n_k} |\bar{n}_1> \ldots |\bar{n}_k>
\nonumber
\eea
\bea
\label{3}
|3> & \sim & (T^{\nu}T^{\mu})_{i_1 j_1}
b^{\mu \dagger}(u)b^{\nu \dagger}(v)|0 > \\
& \otimes &
\epsilon^{i_1 i_2 \ldots i_{N-k}m_1 \ldots m_k} |m_1> \ldots |m_k> \otimes
\epsilon^{j_1 i_2\ldots i_{N-k}n_1 \ldots n_k} |\bar{n}_1> \ldots |\bar{n}_k>
\nonumber
\eea
\bea
\label{4}
|4> & \sim & (T^{\mu})_{i_1 j_1} (T^{\nu})_{i_2 j_2}
b^{\mu \dagger}(u)b^{\nu \dagger}(v)|0 > \\
& \otimes &
\epsilon^{i_1 i_2 i_3 \ldots i_{N-k}m_1 \ldots m_k} |m_1> \ldots |m_k> \otimes
\epsilon^{j_1 j_2 i_3 \ldots i_{N-k}n_1 \ldots n_k} |\bar{n}_1> \ldots |\bar{n}_k>
\nonumber
\eea

In general we are now in a position to calculate the spectrum of
the two particle subspace for general N and $k$ but for simplicity
we will consider only the simplest case of boundary charge $k=1$
where the four state vectors form an overdetermined system.
For SU(2) only two of the states are independent and for SU(N), $N \ge3$
there are only three possible singlet configurations.  These claims
are most clearly verified via the methods of Young tableaux.
Essentially these states correspond to diagonal, anti-symmetric and
symmetric charge combinations where in the case of SU(2) diagonal and
symmetric combinations
are identified. We denote the normalized state vectors
\bea
\label{k1}
|1> &=& \frac{1}{ \sqrt{2 N (N^2 -1)} } ~\delta^{\mu \nu} \delta^{m n}
\label{states} \\
|2> &=& \frac{1}{ \sqrt{ N (N^2 -1)} } ~[T^{\mu}_F, T^{\nu}_F]_{m n }
\nonumber  \\
|3> &=& \sqrt{ \frac{N}{(N^2 -1) (N^2 -4)} }
~( \{T^{\mu}_F, T^{\nu}_F \}_{m n } -
\frac{1}{N} \delta^{\mu \nu} \delta^{m n} )
\nonumber
\eea
where $m=m_1$ and $n=n_1$ in our previous notation.

Considering only particle number conserving terms in a normal ordering
of the Hamiltonian (\ref{hamil}), under Wick's theorem we find a
number of different contributions to the effective theory in the
two particle subspace.
Considerable simplification is achieved though by keeping only terms up to $O(\hbar^0)$ determined via the dimension counting
arguments of \cite{coleman}. Hence,
in operator form, in center of mass coordinates with momentum
$p$ we have, acting on the colour indices, the effective Hamiltonian
\bea
H &= & 4 ( p^2 + m^2)^{1/2} ~\delta^{\lambda \mu} \delta^{\rho \nu}
\delta^{r m} \delta^{s n} + 2 g^2 |x| (T^a_A)_{\lambda \nu}
(T^a_A)_{\mu \rho} ~\delta^{r m} ~\delta^{s n}
\label{effhamil} \\
&& + 2 g^2 x (T^a_A)_{\lambda \mu}  \delta^{\rho \nu}
[ (T^a _F)_{r m } \delta^{s n} +  (T^a _F)_{n s } \delta^{m r}] \nonumber
\eea

In basis of states (\ref{states}) we find the Hamiltonian
(\ref{effhamil}) is a matrix on the
colour charge subspace.  Since we are only interested in the
effect of the Hamiltonian up to $O(\hbar^0)$, we treat the
operators in this matrix as c-numbers and diagonalize.
The result is a set a independent one dimensional quantum mechanical
problems given by the set of Hamiltonians
\beq
\label{Hamiltonian}
H_1 =2(p^2 + m^2)^{1/2}~,~H_2 = 2(p^2 + m^2)^{1/2} + (N+1) g^2 |x|~,~
H_3 = 2(p^2 + m^2)^{1/2} + (N-1) g^2 |x|
\label{su3}
\eeq

For large fermion mass, the classical approximation to
zeroth order in $\hbar$ is both justified and calculable.  In this limit
the solutions of the relevant Schr\"odinger equations are given by
Airy functions and from these solutions we can determine the meson
spectrum in this model.  Without direct calculation though we can make
a few observations. First consider the Schr\"odinger equation
$H \psi =E\psi$ and define string tension here as the
binding energy of the adjoint pair per unit separation
\beq
\sigma_{QM} = \frac{ E- 2m}{|x|}
\label{strten2}
\eeq
In the limit of infinite fermion mass we neglect the momentum in
(\ref{su3}) and we find string tensions $0$, $g^2 (N-1)$ and $g^2 (N+1)$ for
$H_1$, $H_2$ and $H_3$ respectively.
Comparing these results with the string tension as defined
in (\ref{strten}) for adjoint loops in pure Yang-Mills theory with
$k=1$ we find complete agreement.  This establishes a connection between
the group theoretic calculation of (\ref{static}) in the loop picture
of Makeenko and Migdal and the more standard approach of particle field
theory with a  topological term inserted.
More importantly, we can make a one to one identification between
four terms in the loop formula (\ref{static})
and four independent contractions we discussed previously
(\ref{1}-\ref{4}). In the specific  case with $k=1$
 (\ref{su3})  one can understand a connection between towers
of states which
can be derived from Hamiltonians (\ref{Hamiltonian}) and
loop formula (\ref{static}). From this connection it is easy to
interpret the  terms in (\ref{static}) which have  different
string tensions: they simply correspond to the different
basic states (\ref{k1}), each of which interact distinctly with the 
external (topological) sources.

Of note is the state which corresponds to the vanishing
string tension which, in the quantum mechanical problem,  is an
unbound pair of fermions. This pair, through interaction with
external charges, still forms a colour charge singlet but might
be described as free quarks. We already discussed this state
previously with remark that the analogous massless
`` exotic" states appear  in $N$ flavour Schwinger model
for arbitrary mass with specific value of $\theta=\pi k/N$.
In fact these states were noted by Witten long ago  \cite{witten}
and arise for an arbitrary number of adjoint charges in the system.
This is due to the nature of the index structure of the adjoint 
representation which allows the background electric flux string generated
by the boundary charges to be spliced with adjoint charges with the only
energetic cost being the bare mass of the charge.  If we consider a 
single flux string in the system $(k=1)$ with three charges in the 
adjoint representation then the index structure corresponding to  
vanishing string tension is $\bar{T_j} U^j_k V^k_l W^l_m T^m$
(Fig. 1).  Hence we 
see that one line of flux enters the charge on the upper index  and
leaves on the lower so that the string configuration is unchanged.
Additionally,
as long as the charges do not collide we see that their position on the
string is irrelevant and so there is no (long range) force acting between
them and they are unbound.
All of these observations are made for static charges though and
one may expect that for finite fermion mass
exchange interactions may modify this behavior.
However, the $N$ flavour Schwinger model
shows that these states may persist even for finite mass.

\section{Conclusions}

By considering the boundary conditions of two dimensional
QCD with adjoint fermions we have shown  in the limit of
large fermion mass that the multiplicity of vacua
and their physical properties can be
considered in two equivalent ways.  The first involves  considering
the physics in terms of Wilson loops and using the well established
methods of calculation to show the properties of the stable configurations,
including periodicity in the representation of the external loops and
a multiplication of distinct singlet configurations.

Subsequently
we examined the same model in terms of a particle/field theoretic
picture where the boundary conditions were specified by 
static colour charges.
In the limit of large fermion mass we were able to consider the
two particle subspace of the Hamiltonian acting on all possible
singlet configurations of a pair of adjoint fermions and the boundary
charges.  For a particular choice of boundary charge we showed that
the resulting string tensions between the fermions matched exactly
with the calculations for adjoint Wilson loops.

From the field theory approach we were able to find exact solutions
for the meson wave functions in terms of Airy functions.  These
results may be of use in the future for the calculation of the
spectra of these mesons and  fermion condensates in each of
the different vacuum states.
In addition to considering the problem of condensates,
one may try to use our results
with heavy quark mass to generalize the string
picture originally derived   in \cite{grosst} for pure gauge
theory in 2D without dynamical degrees of freedom.
Using heavy quarks we have 
essentially introduced  physical degrees of freedom
without noticeably changing the internal gluodynamics.
Therefore, we expect that the introduction of the heavy
quark might be the first step in the direction
of the string description of   dynamical degrees of freedom.

\section{Acknowledgements}
This work was supported in part by the Natural Sciences and Engineering
Research Council of Canada


\newpage
\section*{Figure Caption}

{\bf Figure 1} Adjoint particles interacting with a single $(k=1)$ string of 
electric flux 

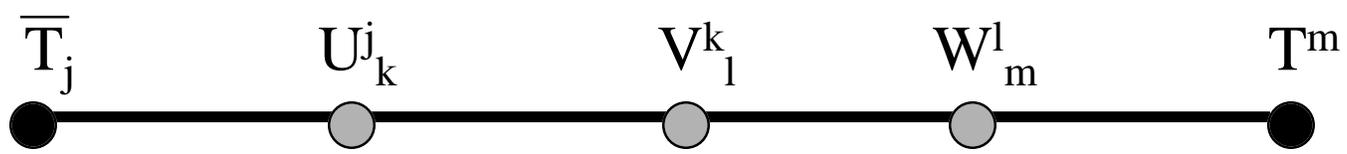
\end{document}